\documentclass[aps,prl,twocolumn,groupedaddress]{revtex4}

\begin{document}

\preprint{hep-ph/0112110 \\
  CERN-TH/2001-334\\
  LMU 01/15}

\title{The Cosmological Evolution of the Nucleon Mass
    and the Electroweak Coupling Constants}


\author{Xavier Calmet}
\email[]{calmet@theorie.physik.uni-muenchen.de}
%\thanks{partially supported by the Deutsche
%Forschungsgemeinschaft, DFG-No. FR 412/27-2}
\affiliation{Ludwig-Maximilians-University Munich, Sektion Physik,
  Theresienstra{\ss}e 37, D-80333 Munich, Germany}
\author{Harald Fritzsch}
%\thanks{partially supported by the VW-Stiftung Hannover (I-77495)}
\affiliation{Stanford Linear Accelerator Center, Stanford University,
  Stanford CA 94309, USA} \affiliation{Ludwig-Maximilians-University
  Munich, Sektion Physik, Theresienstra{\ss}e 37, D-80333 Munich,
  Germany}


\date{\today}

\begin{abstract}
  Starting from astrophysical indications that the fine structure
  constant might undergo a small cosmological time shift, we discuss
  the implications of such an effect from the point of view of
  particle physics.  Grand unification implies small time shifts for
  the nucleon mass, the magnetic moment of the nucleon and the weak
  coupling constant as well. The relative change of the nucleon mass
  is about 40 times larger than the relative change of $\alpha$.
  Laboratory measurements using very advanced methods in quantum
  optics might soon reveal small time shifts of the nucleon mass, the
  magnetic moment of the nucleon and the fine structure constant.
\end{abstract}

%\keywords{}

\maketitle

\section{}
\subsection{}
\subsubsection{}
Some recent astrophysical observations suggest that the fine structure
constant $\alpha$ might change with cosmological time
\cite{Webb:2001mn}. If interpreted in the simplest way, the data
suggest that $\alpha$ was lower in the past:
\begin{equation} \label{exinput}
\Delta \alpha / \alpha = (-0.72 \pm 0.18) \times 10^{-5}
\end{equation}
for a redshift $z \approx 0.5 \ldots 3.5$ \cite{Webb:2001mn}.

The idea that certain fundamental constants might not be constant on a
cosmological time scale was pioneered by Dirac \cite{Dirac}, Milne
\cite{Milne} and P. Jordan \cite{Jordan}. More recently,
time variations of fundamental constants were discussed in connection
to theories based on extra dimensions \cite{Arkani-Hamed:1998rs}.

In this paper we shall study consequences of a possible time
dependence of the fine structure constant, which are expected within
the framework of the Standard Model of the elementary particle
interactions and of unified theories beyond the Standard Model.

In the Standard Model, based on the gauge group $SU(3) \times SU(2)
\times U(1)$, the fine structure constant $\alpha $ is not a basic
parameter of the theory, but is  related to the coupling parameters
$\alpha_i$ ($\alpha_i = g_i^2/(4 \pi)$, where $g_i$ are the coupling
constants of the $SU(3)$, $SU(2)$ or $U(1)$ gauge
interactions.
  
If the three gauge coupling constants are extrapolated to high energy,
they come together at an energy of about $10^{16}$ GeV, as expected,
if the QCD gauge group and the electroweak gauge groups are subgroups
of a simple gauge group, e.g. $SU(5)$ \cite{Georgi:1974sy} or $SO(10)$
\cite{Fritzsch:1975nn}. Thus the scale of the symmetry breaking of the
unifying group determines where the three couplings constants converge
\cite{Georgi:yf}.

If one takes the idea of grand unification seriously, a small shift in
the cosmic time evolution of the electromagnetic coupling constant
$\alpha$ would require that the unified coupling constant
$\alpha_{un}$ undergoes small time changes as well. Otherwise the
grand unification of the three gauge forces would work only at a
particular time. Thus in case of a time dependence one should expect,
that not only the electromagnetic coupling $\alpha$, but all three
gauge couplings $g_1$, $g_2$ and $g_3$ show such a time variation. One
might also consider time changes of other basic parameters, e.g. the
electron mass, but here we shall concentrate on the gauge couplings.

Of special interest is a time variation of the QCD coupling $g_3$.
Taking into account only the lowest order in $\alpha_s=g_3^2/(4 \pi)$,
the behavior of the QCD coupling constant is given by:
\begin{equation}
\alpha_s (\mu ) =
\frac{4 \pi}{\beta_0 \ln \left( \frac{\Lambda^2}{\mu^2} \right)}
\end{equation}
($\mu$: reference scale, $\beta_0 = -11 + \frac{2}{3} n_f, n_f$:
number of quark flavors, $\Lambda $: QCD scale parameter). According
to the experiments one has
$\alpha_s(Q^2=m_Z^2)_{\overline{\mbox{MS}}}=0.1185(20)$.  A typical
value of the scale parameter $\Lambda $ is \cite{Bethke:2000ai}
\begin{equation} \label{ne2}
\Lambda = 213^{+38}_{-35} \mbox{MeV}.
\end{equation}
If $\alpha_s $ is not only a
function of the reference scale $\mu$, but also of the cosmological
time, the scale parameter $\Lambda$ is time-dependent as well.

One finds:
\begin{equation}
\frac{\dot{\alpha }_s}{\alpha_s}
= \frac{2}{\ln \left(\frac{\mu^2}{\Lambda^2} \right)} 
\left( \frac{\dot{\Lambda}}{\Lambda} \right).
\end{equation}
We note that in this relation the coefficient $\beta_0$ has cancelled
out.

The relative changes $\frac{\delta \alpha}{\alpha}$ and $\frac{\delta
  \Lambda}{\Lambda}$ are related by: $\left( \frac{\delta
    \Lambda}{\Lambda} \right) = \left( \frac{\delta
    \alpha_s}{\alpha_s} \right) \ln \left( \mu / \Lambda \right)$.
Thus a relative time shift of $\alpha_s$ (likewise $\alpha_2 $ and
$\alpha_1$) cannot be uniform, i.e. identical for all reference
scales, but changes logarithmically as the scale $\mu $ changes. If
one would identify a relative shift $\left( \delta \alpha_s/ \alpha_s
\right)$ at very high energies, say close to a scale $\Lambda_{G}
\approx 10^{16}$ GeV, given e.g. in a grand unified theory of the
electroweak and strong interactions, the corresponding relative shift
of $\Lambda $ would be larger by a factor $\ln ( \mu / \Lambda )
\approx 38$.

In QCD the proton mass, as well as all other hadronic masses are
proportional to $\Lambda $, if the quark masses are set to zero: $M_p
= {\rm const.}\ \Lambda$. The masses of the light quarks $m_u$, $m_d$
and $m_s $ are small compared to $\Lambda $, however the mass term of
the ``light'' quarks $u$, $d$ and $s$ contributes to the proton mass.
In reality the masses of the light quarks $m_u$, $m_d$ and $m_s$ are
non-zero, but these mass terms contribute only a relatively small
amount (typically less than $10 \%$) to the mass of the nucleon or
nucleus. Here we shall neglect those contributions. The mass of the
nucleon receives also a small contribution from electromagnetism of
the order of $1 \%$, which we shall neglect as well.

If the QCD coupling constant $\alpha _s$ or likewise the QCD scale
parameter $\Lambda $ undergoes a small cosmological time shift, the
nucleon mass as well as the masses of all atomic nuclei would change
in proportion to $\Lambda $. Such a change can be observed by
considering the mass ratio $m_e / m_p$.  Since a change of $\Lambda $
would not affect the electron mass, the electron-proton mass ratio
would change in cosmological time.

The three coupling constants $\alpha_1, \alpha_2$ and $\alpha_s$ seem
to converge, when extrapolated to very high energies, as expected in
grand unified theories. However, in the Standard Model they do not
meet at one point, as expected e.g. in the simplest $SU(5)$-theory
\cite{Georgi:1974sy}.

In models based on the gauge group $SO(10)$ \cite{Fritzsch:1975nn} a
convergence of the three coupling constants can be achieved, if
intermediate energy scales are considered \cite{Chang:uy}. In the
minimal supersymmetric extension of the Standard Model the three gauge
coupling constants do meet at one point \cite{Amaldi:1991cn}.

We consider a theory where the physics affecting the unified coupling
constant is taking place at a scale above that of the unification.
The main assumption is that the physics responsible for a cosmic time
evolution of the coupling constants takes place at energies above the
unification scale. This allows to use the usual relations from grand
unified theories to evolve the unified coupling constant down to low
energy. For example, in string theory the coupling constants are
expectation values of fields.  They might have some cosmological time
evolution \cite{Damour:1994zq}. But, at energies below the grand
unification point, the usual quantum field theory remains valid.

Whatever the correct unification theory might be, one expects in
general that a cosmological time shift affects primarily the unified
single coupling constant $\alpha_{un}$, defined e.g. at the point of
unification. In order to be specific, we shall consider the
supersymmetric $SU(5)$ grand unified theory broken to the gauge group
of the minimal supersymmetric extension of the Standard Model (MSSM)
to derive consequences for low energy physics. As usual the scale for
supersymmetry breaking is assumed to be in the TeV range. However, our
main conclusions will not depend significantly on this assumption.

The scale evolution of the coupling constants in the 1-loop
approximation is given by the well-known relation
\begin{eqnarray}
  \frac{1}{\alpha_i(\mu)}=\frac{1}{\alpha^0_i(\mu^0)}+\frac{1}{2 \pi} b_i \ln
  \left ( \frac{\mu^0}{\mu} \right).
\end{eqnarray}
The parameters $b_i$ are given by $b^{{SM}}_i\!\!\!=(b^{{SM}}_1, b^{{SM}}_2,
b^{{SM}}_3)=(41/10, -19/6, -7)$ below the supersymmetric scale and by
$b^{{S}}_i\!\!\!=(b^{{S}}_1, b^{{S}}_2, b^{{S}}_3)=(33/5, 1,
-3)$ when ${\cal N}=1$ supersymmetry is restored.

Suppose that the coupling constants $\alpha_i$ depend not only on the
scale $\mu$, but also on the cosmological time $t$: $\alpha_i(\mu, t)$.
Since the coefficients $b_i$ are time independent, one finds
\begin{eqnarray} \label{ne1}
\frac{1}{\alpha_i(\mu)}  \frac{\dot\alpha_i(\mu)}{\alpha_i(\mu)}
=
\frac{1}{\alpha_i(\mu')}  \frac{\dot\alpha_i(\mu')}{\alpha_i(\mu')} , \ \ \ i \in \{1,2,3\}
\end{eqnarray}
i.e. the quantity $\alpha_i^{-1} (\dot \alpha_i/\alpha_i)$ is scale
independent.

Since we have to evolve the coupling constants down to energies below
the supersymmetry breaking scale, we have to take into account the
fact that supersymmetry is broken at low energy. We thus have,
replacing the thresholds of the supersymmetric particles by a simple
step function,
\begin{eqnarray} \label{running}
  \alpha_i(\mu)^{-1}
 &=&
  \left ( \frac{1}{\alpha^0_i(\Lambda_{G})}+\frac{1}{2 \pi}
  b^{S}_i  \ln
  \left ( \frac{\Lambda_{G}}{\mu} \right) \right) \theta(\mu- \Lambda_{S})
\\ \nonumber &&  +
\left ( \frac{1}{\alpha^0_i(\Lambda_{S})}+\frac{1}{2 \pi}
  b^{SM}_i   \ln
  \left ( \frac{\Lambda_{S}}{\mu} \right) \right)
\theta(\Lambda_{S}-\mu).
\end{eqnarray}
Here $\Lambda_{S}$ is the supersymmetry breaking scale and
\begin{eqnarray}
  \frac{1}{\alpha^0_i(\Lambda_{S})} &=&
\frac{1}{\alpha^0_i(M_Z)}
  +\frac{1}{2 \pi}
  b^{SM}_i \ln
  \left ( \frac{M_Z}{\Lambda_{S}} \right)
\end{eqnarray}
where $M_Z$ is the $Z$-boson mass and $\alpha^0_i(M_Z)$ is the value
of the coupling constant under consideration measured at $M_Z$.  We
use the following definitions for the coupling constants:
\begin{eqnarray} \label{def1}
  \alpha_1&=& 5/3 g_1^2/(4 \pi)=
  5 \alpha / ( 3 \cos^2(\theta)_{\overline{\mbox{MS}}})
    \\ \nonumber 
  \alpha_2&=& g_2^2/(4 \pi)= \alpha / \sin^2(\theta)_{\overline{\mbox{MS}}}
\\ \nonumber
\alpha_s&=& g_3^2/(4 \pi).
\end{eqnarray}

We suppose that the unified coupling constant $\alpha _{un}$ undergoes
a time shift $\alpha_{un} (\Lambda_{G}) \rightarrow \alpha'_{un}
(\Lambda_{G}): \alpha'_{un} - \alpha_{un} = \delta \alpha_{un}$.
According to (\ref{ne1}) and to the convergence of the three coupling
constants at the unification point $\Lambda_G=1.5 \times 10^{16}$ GeV
with $\alpha_{un}=0.03853$, one finds:
\begin{eqnarray} \label{neq4}
\frac{1}{\alpha_1(\mu)}  \frac{\dot\alpha_1(\mu)}{\alpha_1(\mu)}
=
\frac{1}{\alpha_2(\mu)}  \frac{\dot\alpha_2(\mu)}{\alpha_2(\mu)}
=
\frac{1}{\alpha_s(\mu)}  \frac{\dot\alpha_s(\mu)}{\alpha_s(\mu)}.
\end{eqnarray}
Furthermore one derives from (\ref{def1})
\begin{eqnarray} \label{neq40}
\frac{1}{\alpha_2(\mu)}  \frac{\dot\alpha_2(\mu)}{\alpha_2(\mu)}
=
\frac{3}{8}\frac{1}{\alpha(\mu)}  \frac{\dot\alpha(\mu)}{\alpha(\mu)}
=
\frac{1}{\alpha_s(\mu)}  \frac{\dot\alpha_s(\mu)}{\alpha_s(\mu)}.
\end{eqnarray}
We note that the electroweak mixing angle $\theta$, i.e. the quantity
$\sin^2\theta$, will also be time dependent, but only for
$\mu\neq\Lambda_G$. At $\mu=\Lambda_G$ it is given by the symmetry
value $\sin^2\theta=3/8$. The factor $3/8$ in (\ref{neq40}) arises from
the factor $5 \alpha / ( 3 \cos^2\theta)$ by taking the time dependence
of $\sin^2\theta$ explicitly into account.

Using $\mu=M_Z$ as the scale parameter in (\ref{ne2}), we obtain at
$\mu=M_Z$, using $\alpha_s(M_Z)=0.121$ \cite{Groom:in}:
\begin{eqnarray} \label{neq3}
\frac{\dot\alpha}{\alpha}= \frac{8}{3}
\frac{\alpha}{\alpha_s} \frac{\dot\alpha_s(\mu)}{\alpha_s(\mu)}
=\frac{8}{3} \frac{\alpha}{\alpha_s}
\frac{1}{\ln \left( \frac{\mu}{\Lambda} \right)} \frac{\dot{\Lambda}}{\Lambda}
\approx 0.0285 \cdot  \frac{\dot{\Lambda}}{\Lambda}.
\end{eqnarray}
Using the scale invariance of $\alpha^{-1} {\dot \alpha}/\alpha$, we
obtain
\begin{eqnarray} 
\frac{\dot\alpha}{\alpha}(\mu=0)&=&\frac{\dot\alpha}{\alpha}(\mu=M_Z) \ 
\frac{\alpha(\mu=0)}{\alpha(M_Z)}
\\ \nonumber
&\approx& 0.93 \cdot \frac{\dot\alpha}{\alpha}(\mu=M_Z).
\end{eqnarray}

The result is:
\begin{eqnarray} \label{result} 
\frac{\dot{\Lambda}}{\Lambda} = R \frac{\dot\alpha}{\alpha}(\mu=0)
\end{eqnarray}
the coefficient $R$ is calculated to $R=37.7\pm 2.3$. The uncertainty
of $R$ is given, according to (\ref{neq3}), by the uncertainty of the
ratio $\alpha/\alpha_s$, which is dominated by the uncertainty of
$\alpha_s$.

We should like to emphasize that the relation (\ref{result}) is
independent of the details of the evolution of the coupling constants
at very high energies, in particular it is independent of the details
of supersymmetry breaking. The Landau pole of (\ref{running}) for $i=3$
corresponds to
\begin{eqnarray}
  \Lambda'&=&\Lambda_{S}
  \exp \left(\frac{2 \pi}{b^{SM}_3} \frac{1}{\alpha_{un}'} \right)
  \left ( \frac{\Lambda_{G}}{\Lambda_{S}}
  \right)^{\left(\frac{b_3^{S}}{b_3^{SM}} \right)}.
\end{eqnarray}
 We find
\begin{eqnarray} \label{re}
\frac{\dot{\Lambda}}{\Lambda}=-\frac{3}{8}\frac{2 \pi}{b_3^{SM}}\frac{1}{\alpha}\frac{\dot{\alpha}}{\alpha}.
\end{eqnarray}
i.e. there is no dependence on $\Lambda_S$. If we calculate
$\dot{\Lambda}/\Lambda$ using the relation above in the case of 6 quark
flavors, neglecting the masses of the quarks, we find $R \approx 46$.

This shows that the actual value of $R$ is sensitive to the inclusion
of the quark masses and the associated thresholds, just like in the
determination of $\Lambda$. Furthermore higher order terms in the QCD
evolution of $\alpha_s$ will play a role. For this reason the
systematic uncertainty in the value of $R$ is certainly larger than
the error given above. We estimate:
\begin{eqnarray} \label{resE}
R=38\pm 6
\end{eqnarray}
taking into account both the experimental error in the determination
of $\alpha_s(M_Z)$ and the systematic uncertainties.

The time change of $\Lambda$ implies a time change of the proton mass
and of all nuclear mass scales, as well as of the pion mass, which
would change in proportion to $\Lambda^{1/2}$, according to the chiral
symmetry realtion $M_\pi^2=\mbox{const.} m_q \Lambda$ ($m_q$: light
quark mass average). We obtain
\begin{eqnarray} \label{result2}
\frac{\dot{M}}{M} = \frac{\dot{\Lambda}}{\Lambda} = R \frac{\dot{\alpha}}{\alpha} \approx 38 \cdot \frac{\dot{\alpha}}{\alpha}.
\end{eqnarray}

Thus the change of the nucleus mass amounts to about 0.3 MeV, if we
base our calculations on the time shift of $\alpha$ given in ref.
\cite{Webb:2001mn}. At a redshift of about one the mass of the nucleon
as well as the masses of the nuclei were about 0.3 ${}^0\!/_{\!00}$
smaller than today.

In QCD the magnetic moment of the proton $\mu = g_p \cdot e / 2M_p$ is
related to the magnetic moments of the constituent quarks. Although it
is not possible to calculate the magnetic moment of the proton with
high precision, the moment scales in proportion to $\Lambda^{-1}$ in
the chiral limit where the quark masses vanish. Thus, we have
\begin{equation} \label{magmom}
\frac{\dot{\mu}_p}{\mu_p}=-\frac{\dot{\Lambda}}{\Lambda} =
-R \frac{\dot{\alpha}}{\alpha}.
\end{equation}

The gyromagnetic ratio $g_p$ will not be time dependent, since the
proton mass scales like $\mu_p^{-1}$, however the ratio of the
magnetic moments $\mu_p/\mu_e$ will be time-dependent:
\begin{equation} 
\dot{\left( \frac{\mu_p}{\mu_e} \right)}/\left( \frac{\mu_p}{\mu_e} \right)
=-\frac{\dot{\Lambda}}{\Lambda}= - R \frac{\dot{\alpha}}{\alpha}.
\end{equation}

The present astrophysical limit on the proton-electron mass ratio
 $\mu= M_p/m_e$ obtained at a redshift of $z=2.81$ is
 \cite{Potekhin:1998mf}
 \begin{eqnarray} \label{eq29}
  -1.7 \times 10^{-5}<\frac{{\Delta \mu}}{\mu} &<& 2\times 10^{-4}.
 \end{eqnarray}
Using (\ref{result2}) and (\ref{exinput}), one would expect:
\begin{eqnarray}
  \frac{\Delta \mu}{\mu} \approx -3 \cdot 10^{-4} 
\end{eqnarray}
a result, which violates the bound (\ref{eq29}), but in view of the
large errors on the astrophysical side we do not regard this as a
serious disagreement, rather as a sign that astrophysical data might
soon clarify whether a time change of the nuclear mass scale following
(\ref{result2}) is indeed present.
% %Again we obtain a disagreement with our result (\ref{Nucmass}) obtained from
% %eq. (\ref{expconstr})

% {\bf Either the data on a time variation of $\alpha$ must be wrong, or the
% errors in (\ref {expconstr}) are largely underestimated, as also
% mentioned in ref. \cite{Murphy:2001nu}, or the idea of a grand
% unification of the gauge coupling constants, as described above, is
% not correct.}

A clarification of the situation could come from laboratory
experiments. Assuming an age of the universe of the order 14 Gyr, the
various astrophysical limit can be used to derive relative changes of
the various quantities, e.g. $\dot{\alpha}/\alpha$ or
$\dot{\Lambda}/\Lambda$, per year, assuming for simplicity a linear
time evolution. The constraint on $|\Delta \mu_p/\mu_p|$ given above
(\ref{eq29}) leads to \cite{Potekhin:1998mf}:
\begin{eqnarray}
  \left| \frac{{\dot \mu}_p}{\mu_p}\right|&<& 1.5 \times 10^{-14}
 \ \mbox{yr$^{-1}$}.
\end{eqnarray}
% The corresponding limits from astrophysics on $\dot{\alpha}/\alpha$
% and $\dot{g}_p/g_p$ are \cite{Drinkwater:1997qj}:
% \begin{eqnarray}
%  \left |\frac{ \dot{\alpha}}{\alpha}\right | < 1 \times 10^{-15} \
% \mbox{yr$^{-1}$} \ \mbox{at} \ \ z=0.25  \ \ \mbox{and}  \ \ 
% \left |\frac{\dot{g}_p}{g_p}\right |< 2 \times 10^{-15} \ \mbox{yr$^{-1}$}. 
% \end{eqnarray}

Direct laboratory measurements provide the constraint \cite{Prestage}:
\begin{eqnarray}
\left | \frac{\dot{\alpha}}{\alpha}\right |&\le& 3.7 \times 10^{-14} \ \mbox{yr$^{-1}$}.
\end{eqnarray}

Using advanced methods in quantum optics, it seems possible to improve
the present laboratory limits for a time variation of $\alpha$ and of
the nucleon mass by several orders of magnitude. A time variation of
$\alpha$ could be observed by monitoring the atomic fine structure in
a period of several years. Monitoring the rotational and/or
vibrational transition frequencies of molecules, e.g. diatomic
molecules like H$_2$ or CO would allow to set stringent limits on a
time variation of the nucleon mass.  

According to our estimates, the largest effect is expected to be a
cosmological time shift of the nucleon mass, observed e.g. by
monitoring molecular frequencies. Due to the relation (\ref{magmom})
similar effects (same amounts, opposite sign) should be seen in a time
shift of $\mu_p$, observed by monitoring hyperfine transitions. These
effects should be about 40 times larger than a time shift of $\alpha$
(see eq.(\ref{result})), observed e.g. in monitoring fine structure
effects. In quantum optics one may achieve a relative accuracy in
frequency measurements of the order of $\Delta \omega/\omega\approx
10^{-18}$, which would allow to improve the present limits
significantly or observe effects of time variation. We note, however
that the present continuously operated atomic frequency standards (H,
Cs, Hg$^+$) are using transitions between ground states hyperfine
energy levels, given by the interaction of a nuclear magnetic moment
with the magnetic moment of the valence electron \cite{Prestage}. In a
relative comparison the time dependence of the nuclear magnetic
moments drops out. In order to see an effect, following
(\ref{magmom}), a comparison with a frequency standard independent of
the nuclear magnetic moments is necessary.

It is quite possible that future laboratory experiments find positive
effects for time variations of $M_p$, $\mu_p$ and $\alpha$.  If a time
variation is observed, the actual amount of time variation, say the
value of $\dot{M}/M$, would be an important parameter to connect
particle physics quantities with the cosmological evolution.

Finally we should like to mention that the link between the various
coupling constants of the Standard Model discussed here implies that
nuclear physics scales, including the pion mass, change as well. For
this reason the constraints on a time variation of $\alpha$ derived
from an analysis of the natural reactor at Oklo (Gabon, Africa)
\cite{OKLO} cannot be taken seriously.  In fact, it is a bound on the
product $\alpha M_\pi$ under the additional assumption that other
nuclear physics and strong interaction parameters do not change. The
product $\alpha M_\pi$ would change, according to the relation
(\ref{result}) as $\dot{\alpha}/\alpha+\dot{\Lambda}/(2\Lambda)
\approx 21 \dot{\alpha}/\alpha$, since $M_\pi$ is proportional to
$\sqrt{m \Lambda}$ ($m$: light quark mass). This would lead to a bound
about an order of magnitude stronger than the present bound on the
time variation of $\alpha$. However other nuclear physics parameters,
change as well. A more detailed analysis of the nuclear physics
aspects of a time change of $\Lambda$ is needed in order to see
whether there is a disagreement here.

Furthermore we expect a small cosmological time shift of the $n-p$
mass difference. This would affect the cosmic nucleosynthesis of the
light elements. An analysis of nucleosynthesis will be made elsewhere.

\begin{acknowledgments}
We are grateful to T.~Haensch and to H.~Walther for discussions on the
potential of the measurements of the time variation of $\alpha$ and of
the nucleon mass using quantum optics methods.  Furthermore we are
indebted to E.~Kolb, A.~Hebecker, J.~March-Russell, P.~Minkowski and
G.~Steigman for discussions and to T.~Dent for a useful communication.
\end{acknowledgments}



\begin{thebibliography}{10}  
%%%%%%%%%%%%%%%%%%%%%%%%%%%%%%%%%%%%%%%%%%%%%%%%%%%%%%%%%%%%%%%%%%%%%% 
%\cite{Webb:2001mn}
\bibitem{Webb:2001mn}
J.~K.~Webb {\it et al.},
%``Further Evidence for Cosmological Evolution of the Fine Structure Constant,''
Phys.\ Rev.\ Lett.\  {\bf 87} 091301 (2001) 
[arXiv:astro-ph/0012539].
%%CITATION = ASTRO-PH 0012539;%%
%%%%%%%%%%%%%%%%%%%%%%%%%%%%%%%%%%%%%%%%%%%%%%%%%%%%%%%%%%%%%%%%%%%%%%%%%%%%%
\bibitem{Dirac}
P.~M.~Dirac, Nature {\bf192}, 235 (1937).
\bibitem{Milne}
E.~A.~Milne, Relativity, Gravitation and World Structure, Clarendon press, Oxford, (1935), Proc. Roy. Soc. A, 3, 242 (1937).
\bibitem{Jordan}
P.~Jordan, Naturwiss., 25, 513 (1937), Z. Physik, 113, 660 (1939).


%%%%%%%%%%%%%%%%%%%%%%%%%%%%%%%%%%%%%%%%%%%%%%%%%%%%%%%%%%%%%%%%%%%%%%%%%%%%%


\bibitem{Arkani-Hamed:1998rs}
%\cite{Marciano:1983wy}
%\bibitem{Marciano:1983wy}
W.~J.~Marciano,
%``Time Variation Of The Fundamental 'Constants' And Kaluza-Klein Theories,''
Phys.\ Rev.\ Lett.\  {\bf 52}, 489 (1984),
%%CITATION = PRLTA,52,489;%%
%\cite{Dvali:2001dd}
%\bibitem{Dvali:2001dd}
G.~R.~Dvali and M.~Zaldarriaga,
%``Changing alpha with time: Implications for fifth-force-type experiments  and quintessence,''
arXiv:hep-ph/0108217,
%%CITATION = HEP-PH 0108217;%%
T.~Banks, M.~Dine, M.~ R.~Douglas, hep-ph/0112059.
%%%%%%%%%%%%%
%\cite{Georgi:1974sy}
\bibitem{Georgi:1974sy}
H.~Georgi and S.~L.~Glashow,
%``Unity Of All Elementary Particle Forces,''
Phys.\ Rev.\ Lett.\  {\bf 32}, 438 (1974).
%%CITATION = PRLTA,32,438;%%
%%%%%%%%%%%%%
%\cite{Fritzsch:1975nn}
\bibitem{Fritzsch:1975nn}
H.~Fritzsch and P.~Minkowski,
%``Unified Interactions Of Leptons And Hadrons,''
Annals Phys.\  {\bf 93}, 193 (1975),
H.~Georgi, in {\it Particles and Fields}, (AIP, New York, 1975).
%%%%%%%%%%%%%%%
%\cite{Georgi:yf}
\bibitem{Georgi:yf}
H.~Georgi, H.~R.~Quinn and S.~Weinberg,
%``Hierarchy Of Interactions In Unified Gauge Theories,''
Phys.\ Rev.\ Lett.\  {\bf 33}, 451 (1974).
%%CITATION = PRLTA,33,451;%%
%%%%%%%%%%%%%%%
%%CITATION = APNYA,93,193;%%%%%%%%%%%%%%%
%\cite{Bethke:2000ai}
\bibitem{Bethke:2000ai}
S.~Bethke,
%``Determination of the QCD coupling alpha(s),''
J.\ Phys.\ G {\bf 26}, R27 (2000)
[arXiv:hep-ex/0004021].
%%CITATION = HEP-EX 0004021;%%
%%%%%%%%%%%%%




%%%%%%%%%%%%%
%\cite{Chang:uy}
\bibitem{Chang:uy}
D.~Chang, R.~N.~Mohapatra and M.~K.~Parida,
%``A New Approach To Left-Right Symmetry Breaking In Unified Gauge Theories,''
Phys.\ Rev.\ D {\bf 30}, 1052 (1984),
%%CITATION = PHRVA,D30,1052;%%
%\cite{Mohapatra:1992dx}
%\bibitem{Mohapatra:1992dx}
R.~N.~Mohapatra and M.~K.~Parida,
%``Threshold effects on the mass scale predictions in SO(10) models and solar neutrino puzzle,''
Phys.\ Rev.\ D {\bf 47}, 264 (1993)
[arXiv:hep-ph/9204234].
%%CITATION = HEP-PH 9204234;%%




%%%%%%%%%%%%%
%\cite{Amaldi:1991cn}
\bibitem{Amaldi:1991cn}
U.~Amaldi, W.~de Boer and H.~Furstenau,
%``Comparison of grand unified theories with electroweak and strong coupling constants measured at LEP,''
Phys.\ Lett.\ B {\bf 260}, 447 (1991).
%%CITATION = PHLTA,B260,447;%%
%%%%%%%%%%%%%

%%%%%%%%%%%%%
%\cite{Damour:1994zq}
\bibitem{Damour:1994zq}
  %\cite{Green:sp}
%\bibitem{Green:sp}
M.~B.~Green, J.~H.~Schwarz and E.~Witten, Superstring Theory, Vol. 1 and 2,
T.~Damour and A.~M.~Polyakov,
%``The String dilaton and a least coupling principle,''
Nucl.\ Phys.\ B {\bf 423}, 532 (1994)
[arXiv:hep-th/9401069].
%%CITATION = HEP-TH 9401069;%%
%%%%%%%%%%%%%


%%%%%%%%%%%%%
%\cite{Groom:in}
\bibitem{Groom:in}
D.~E.~Groom {\it et al.}  [Particle Data Group Collaboration],
%``Review Of Particle Physics,''
Eur.\ Phys.\ J.\ C {\bf 15}, 1 (2000).
%%CITATION = EPHJA,C15,1;%%
%%%%%%%%%%%%%

%% %%%%%%%%%%%%
% %\cite{Murphy:2001nu}
% \bibitem{Murphy:2001nu}
% M.~T.~Murphy, J.~K.~Webb, V.~V.~Flambaum, M.~J.~Drinkwater, F.~Combes and T.~Wiklind,
% %``Improved constraints on possible variation of physical constants from H I 21cm and molecular QSO absorption lines,''
% Mon.\ Not.\ Roy.\ Astron.\ Soc.\  {\bf 327}, 1244 (2001)
% [arXiv:astro-ph/0101519].
% %%CITATION = ASTRO-PH 0101519;%%
% %%%%%%%%%%%%%

%%%%%%%%%%%%%
%\cite{Potekhin:1998mf}
\bibitem{Potekhin:1998mf}
A.~Y.~Potekhin, A.~V.~Ivanchik, D.~A.~Varshalovich, K.~M.~Lanzetta, J.~A.~Baldwin, G.~M.~Williger and R.~F.~Carswell,
%``Testing cosmological variability of the proton-to-electron mass ratio  using the spectrum of PKS 0528-250,''
Astrophys.\ J.\  {\bf 505}, 523 (1998)
[arXiv:astro-ph/9804116].
%%CITATION = ASTRO-PH 9804116;%%
%%%%%%%%%%%%%
%%%%%%%%%%%%%
%\cite{Drinkwater:1997qj}
%\bibitem{Drinkwater:1997qj}
%M.~J.~Drinkwater, J.~K.~Webb, J.~D.~Barrow and V.~V.~Flambaum,
%``Limits on the Variability of Physical Constants,''
%arXiv:astro-ph/9709227.
%%CITATION = ASTRO-PH 9709227;%%
%%%%%%%%%%%%%
%%%%%%%%%%%%%
%\cite{Prestage}
\bibitem{Prestage}
J.~D.~Prestage, L.~.T.~Robert, and L.~Maleki,
Phys.\ Rev.\ Lett.\  {\bf 74} 3511 (1995).
%%%%%%%%%%%%%
%%%%%%%%%%%%%
\bibitem{OKLO}
  A.~I.~ Shylakhter, Nature, {\bf 264}, 340, 1976; ATOMKI Report A/l, 1983.
%%%%%%%%%%%%%
\end{thebibliography}
\end{document}